\documentclass{PoS}

\usepackage{graphicx}

\usepackage{subfigure}
\usepackage{amsmath}

\usepackage{xspace}

\newcommand{\muT}{\left(\frac{\mu}{T}\right)}

\newcommand{\twofigs}{0.44\linewidth}
\newcommand{\threefigs}{0.32\linewidth}
\newcommand{\Phibar}{\ensuremath{\bar{\Phi}}}
\newcommand{\LPQM}{\ensuremath{\mathcal{L}_{\textrm{PQM}}}\xspace}
\newcommand{\Dslash}{\ensuremath{D\hspace{-1.5ex} /}}
\newcommand{\Tr}{\ensuremath{\operatorname{Tr}}}

\newcommand{\vev}[1]{\ensuremath{\left\langle #1 \right\rangle}}
\newcommand{\einh}[1]{\ensuremath{\,\text{#1}}}
\newcommand{\MeV}{\einh{MeV}}

\def\Fig#1{Fig.~\ref{#1}}

\graphicspath{
{./figures/}
}

\title{QCD thermodynamics with effective models}

\ShortTitle{QCD thermodynamics with effective models}
\author{\speaker{Bernd-Jochen
    Schaefer}\\
  Institut f\"ur Physik, Karl-Franzens-Universit\"at, A-8010
  Graz, Austria\\
  E-mail: \email{bernd-jochen.schaefer@uni-graz.at}}

\author{Mathias Wagner\\
        Institut f\"ur Kernphysik, TU Darmstadt, D-64289 Darmstadt, Germany and\\
        ExtreMe Matter Institute EMMI,  GSI Helmholtzzentrum f\"{u}r Schwerionenforschung GmbH, D-64291 Darmstadt, Germany\\
        E-mail: \email{mathias.wagner@physik.tu-darmstadt.de}}

\author{Jochen Wambach\\
        Institut f\"ur Kernphysik, TU Darmstadt, D-64289 Darmstadt, Germany and\\
        Gesellschaft f\"{u}r
  Schwerionenforschung GSI, D-64291 Darmstadt, Germany\\
        E-mail: \email{jochen.wambach@physik.tu-darmstadt.de}}

      \abstract{In this talk we extend the Polyakov-quark-meson model
        to $N_f=2+1$ quark flavors and study its bulk thermodynamics
        at finite temperatures in mean-field approximation. Three
        different Polyakov-loop potentials are considered. Our
        findings are confronted to recent QCD lattice simulations of
        the RBC-Bielefeld and HotQCD collaborations. Furthermore, the
        finite chemical potential expansion of the quark-number
        susceptibility in a Taylor series around vanishing chemical
        potential is analyzed. By means of a novel algorithmic
        differentiation technique, we have calculated Taylor
        coefficients up to 24$^\textrm{th}$ order in the model for the
        first time. This allows the systematic study of convergence
        properties of the Taylor series. }

\FullConference{5th International Workshop on Critical Point and Onset of Deconfinement - CPOD 2009,\\
		 June 08 - 12 2009\\
		 Brookhaven National Laboratory, Long Island, New York, USA}

\begin{document}

\section{Introduction}

The deeper understanding of strongly interacting matter under extreme
conditions plays a crucial role in many research programs. In
particular, the search for possible (tri)critical endpoints in the
phase diagram of strongly interacting matter is a major focus of,
e.g., the planned CBM experiment at the FAIR facility. The underlying
theory, QCD predicts several phases which are linked to certain phase
transitions. For example, the confined, hadronic phase at low
temperatures and chemical potentials (net quark densities) is
separated from the deconfined quark-gluon plasma phase at high
temperature and densities. Chiral and deconfinement aspects of these
QCD phase transitions are the major focus of this talk.

For small temperatures and finite chemical potentials the chiral phase
transition is probably of first-order while a crossover is expected at
high temperatures and small chemical potentials. This suggests the
existence of at least one critical endpoint (CEP) where the
first-order transition line in the phase diagram terminates. At the
CEP the transition is of second-order. The location and even the
possible existence of additional endpoints in the phase diagram is
still an open question (see e.g.\cite{Stephanov:2007fk}). Furthermore,
the coincidence of the chiral and the deconfinement transition at
vanishing chemical potentials and a possible confined but chirally
symmetric phase, the quarkyonic phase, \cite{McLerran:2007qj} at
finite chemical potentials are much under debate.

Different regimes of the QCD phase diagram can be explored by
employing various theoretical methods. Lattice QCD simulations are
applicable at zero or imaginary chemical potentials but at finite real
chemical potentials the fermion sign problem is still a considerable
obstacle. At finite temperatures recent lattice simulations describe
the QCD thermodynamics very trustfully since larger volumes and quark
masses closer to their physical values are used. Despite this progress
some lattice groups differ in their predictions even at vanishing
chemical potentials. The RBC-Bielefeld group sees a coincidence of the
chiral and deconfinement transition while the Wuppertal group obtains
a larger critical temperature for the deconfinement
transition~\cite{Cheng, Bazavov:2009zn, Aoki}. Several extrapolations
techniques towards small finite chemical potentials such as the
reweighting method, imaginary chemical potential or a Taylor expansion
around vanishing chemical potentials have been proposed (see
\cite{SchmidtPhilipsen} for an overview).

Another theoretical method is based on NJL-type or quark-meson models
which incorporate chiral symmetry breaking in an effective manner.
These models do not suffer from any limitations at finite chemical
potentials or finite volumes but cannot address the deconfinement
transition due to the lack of confinement. However, such types of
models can be augmented with the Polyakov loop resulting in
Polyakov-loop NJL-type (PNJL) or Polyakov-loop quark-meson (PQM)
models. With these extended models both phase transitions become
accessible.

In this talk we report on recent results from a chiral PQM model with
$N_f=2+1$ quark flavors. We compare the bulk thermodynamics of this
model for various Polyakov-loop potentials with newer QCD lattice data
where the larger quark masses on the lattice are also explicitly
considered in the model comparison.

Based on a novel numerical method, the algorithmic differentiation
technique, higher derivatives of the thermodynamic potential can be
extremely precise calculated \cite{ADTAylor}. This method, applied to
the PQM model, allows us to investigate convergence properties of the
Taylor expansion method used on the lattice.

\section{Polyakov-quark-meson model for three quark flavors}

The Polyakov-quark-meson model for three quark flavors is a
combination of the chiral linear $\sigma$-model with the Polyakov loop
$\Phi(\vec x)$, the thermal expectation value of a color traced Wilson
loop in temporal direction \cite{Schaefer:2008ax}. Since the Polyakov
loop in these types of effective models is used as a classical
variable, implementation details of this quantity are not important.
It serves as an order parameter of the center symmetry in the limit of
infinitely heavy quarks. It is finite at high temperatures
corresponding to the deconfined plasma phase, where the center
symmetry is spontaneously broken and it vanishes in the confined,
center-symmetric phase. However, in a system with dynamical quarks the
center symmetry is always broken explicitly but $\Phi$ still seems to
be a useful indicator of the confinement/deconfinement transition. The
PQM Lagrangian consists of a quark-meson part and a Polyakov-loop
potential $\mathcal{U} (\Phi, \Phibar)$, which depends on the
Polyakov-loop variable $\Phi$ and its hermitian conjugate $\Phibar$.
The coupling of the background gauge field to the quarks is achieved
by replacing the standard derivative $\partial_\mu$ in the quark-meson
contribution with a covariant derivative $D_\mu = \partial_\mu - i
A_\mu$ with $A_\mu = \delta_{\mu 0} A^0$. This leads to the Lagrangian
\begin{equation}
  \label{eq:lpqm}
  \LPQM = \bar{q}\left(i \Dslash - g \phi_5 \right) q +
  \mathcal{L}_m  
  -\mathcal{U} (\Phi, \Phibar)\ ,
\end{equation}
where the interaction between the scalar ($\sigma_a$) and the
pseudoscalar ($\pi_a$) meson nonets and the three quark flavors $q$ is
implemented by an Yukawa-type vertex
\begin{equation}
  \phi_5 = \frac{ \lambda_a}{2}\left( \sigma_a + i \gamma_5 \pi_a
  \right) \ .
\end{equation}
The remaining, purely mesonic contribution reads
\begin{eqnarray}
\label{eq:mesonL}
  \mathcal{L}_m &=& \Tr \left( \partial_\mu \phi^\dagger \partial^\mu
    \phi \right)
  - m^2 \Tr ( \phi^\dagger \phi) -\lambda_1 \left[\Tr (\phi^\dagger
    \phi)\right]^2 
  - \lambda_2 \Tr\left(\phi^\dagger \phi\right)^2\nonumber \\
  &&  +c   \left(\det (\phi) + \det (\phi^\dagger) \right)
   + \Tr\left[H(\phi+\phi^\dagger)\right]\ ,
\end{eqnarray}
with the short-hand notation
$\phi = \lambda_a \left(\sigma_a + i \pi_a\right)/2$. The Polyakov
loop potential $\mathcal{U}$ preserves the center symmetry of the pure
Yang-Mills (YM) theory and several explicit choices are possible which
we compare with each other in the following.

\subsection{Different choices of the Polyakov loop potential}

Several parameterizations of the Polaykov loop potential have been
proposed in the last years, see \cite{poly}. The simplest choice is
based on a Ginzburg-Landau ansatz and results in an expansion in terms
of the order parameter
\begin{equation}
\label{eq:upoly}
  \frac{\mathcal{U}_{\text{poly}}}{T^{4}}= -b_2
\left(|\Phi|^2+|\bar\Phi|^2 \right) 
-b_3(\Phi^3+\Phibar^3)+b_4
\left(|\Phi|^2+|\Phibar|^2\right)^2
\end{equation}
with a temperature-dependent coefficient
$b_2(T) = a_0 + a_1 \left(\frac{T_0}{T}\right) + a_2
\left(\frac{T_0}{T}\right)^2 + a_3 \left(\frac{T_0}{T}\right)^3$. The
parameters are adjusted to the pure gauge lattice data. An improved
ansatz, motivated by the $SU(3)$ Haar measure, results in
\begin{equation}
\label{eq:ulog}
\frac{\mathcal{U}_{\text{log}}}{T^{4}}= -\frac{1}{2}a(T) \Phibar \Phi
+ b(T) \ln \left[1-6 \Phibar\Phi + 4\left(\Phi^{3}+\Phibar^{3}\right)
  - 3 \left(\Phibar \Phi\right)^{2}\right]\ ,
\end{equation}
with the temperature-dependent pre-factors
$ a(T) = a_0 + a_1 \left(\frac{T_0}{T}\right) + a_2
\left(\frac{T_0}{T}\right)^2$ and
$ b(T) = b_3 \left(\frac{T_0}{T}\right)^3$. In both versions the
parameter $T_0 = 270$ MeV corresponds to the transition temperature in
the pure YM theory. The third version, proposed by Fukushima, has only two parameters $a$ and $b$ and is
inspired from a strong-coupling analysis
\begin{equation}
\label{eq:ufuku}
\frac{\mathcal{U}_{\text{Fuku}}}{T^4} = -\frac{b}{T^3}  \left[54
  e^{-a/T} \Phi \Phibar  + \ln \left(1 - 6 \Phi \Phibar - 3 (\Phi
    \Phibar)^2 + 4 (\Phi^3 
    + \Phibar^3)\right)\right]\ .
\end{equation}
The parameter $a$ determines the deconfinement transition in pure
gauge theory, while $b$ controls the mixing of the chiral and the
deconfinement transition. Here, they are not fitted to the lattice
data but also reproduce a first-order phase transition at $T_0 \sim
270$ MeV in the pure YM theory. Details concerning the PQM model can
be found in the forthcoming work \cite{BJSMW2009}.

When dynamical quarks are present in the system the running coupling
of QCD is modified by fermionic contributions. The size of this effect
can be estimated within perturbation theory, see e.g.~\cite{Braun}. As shown
in Ref.~\cite{Schaefer:2007pw} this leads to an $N_f$-modification of
the expansion coefficients in the polynomial Polyakov-loop potential
and can further be mapped onto an $N_f$-dependent $T_0$. In
Tab.~\ref{tab:ptNf} one sees that the critical temperature $T_0$
decreases for increasing $N_f$.
\begin{table}[thbp]
\centering
\begin{tabular}{|c||c|c|c|c|c|c|}
\hline $N_f$ & 0 & 1 & 2 &2+1 & 3\\\hline
$T_0\  [\MeV]$ & 270 & 240 & 208 & 187 & 178\\\hline
\end{tabular}
\caption{\label{tab:ptNf}The critical temperature $T_0$ for $N_f$
  massless flavors according to~\cite{Schaefer:2007pw}. The value for
  $2+1$ flavors has been estimated by using HTL/HDL theory for a
  massive strange quark with $m_s=150 \MeV$.}  
\end{table}

\subsection{Thermodynamic potential}
The thermodynamic potential of the PQM model is evaluated in
mean-field approximation similar to \cite{Schaefer:2007pw,
  Scavenius:2000qd}. It splits into three contributions: the mesonic,
$U\left(\sigma_{x},\sigma_{y}\right)$, the quark/antiquark,
$\Omega_{\bar{q}{q}}$, and the Polyakov loop contributions
\begin{equation}
  \label{eq:grandpot}
  \Omega = U \left(\sigma_{x},\sigma_{y}\right) +
  \Omega_{\bar{q}{q}} \left(\sigma_{x},\sigma_{y}, \Phi,\Phibar \right) +
  \mathcal{U}\left(\Phi,\Phibar\right)\ . 
\end{equation}
In general, the grand potential is a function of the temperature and
three quark chemical potentials, one for each flavor. Here, we
consider the isospin-symmetric case with two degenerated light quark
masses. As a consequence, only two independent quark chemical
potentials, the light $\mu_q$ and the strange $\mu_s$, emerge.
Furthermore, only two order parameters in the meson sector, the
non-strange $\sigma_x$ and strange $\sigma_y$ condensate are present.
Note, the Polyakov loop variables are coupled to the fermionic part
that also depends on the mesonic condensates via the quark masses.
Finally, the temperature and quark chemical potential dependence of
all four order parameters for the chiral and deconfinement transition
are determined as solutions of the corresponding coupled equations of
motion, i.e.,
\begin{equation}
  \label{eq:pqmeom}
  \left.\frac{ \partial \Omega}{\partial 
      \sigma_x} = \frac{ \partial \Omega}{\partial \sigma_y}  = \frac{
      \partial \Omega}{\partial \Phi}  =\frac{ \partial
      \Omega}{\partial \Phibar} 
  \right|_{\text{min}} = 0\ ,
\end{equation}
where
$\text{min}=\left(\sigma_x=\vev{\sigma_x}, \sigma_y=\vev{\sigma_y},
  \Phi=\vev{\Phi}, \bar\Phi=\vev{\bar\Phi} \right)$ labels the global
minimum of the grand potential.

\begin{figure*}[htbp]
  \centering \subfigure[$\ $Polyakov loop expectation value
  $\vev \Phi$]{\label{sfig:pqmpoly}
    \includegraphics[width=\twofigs]{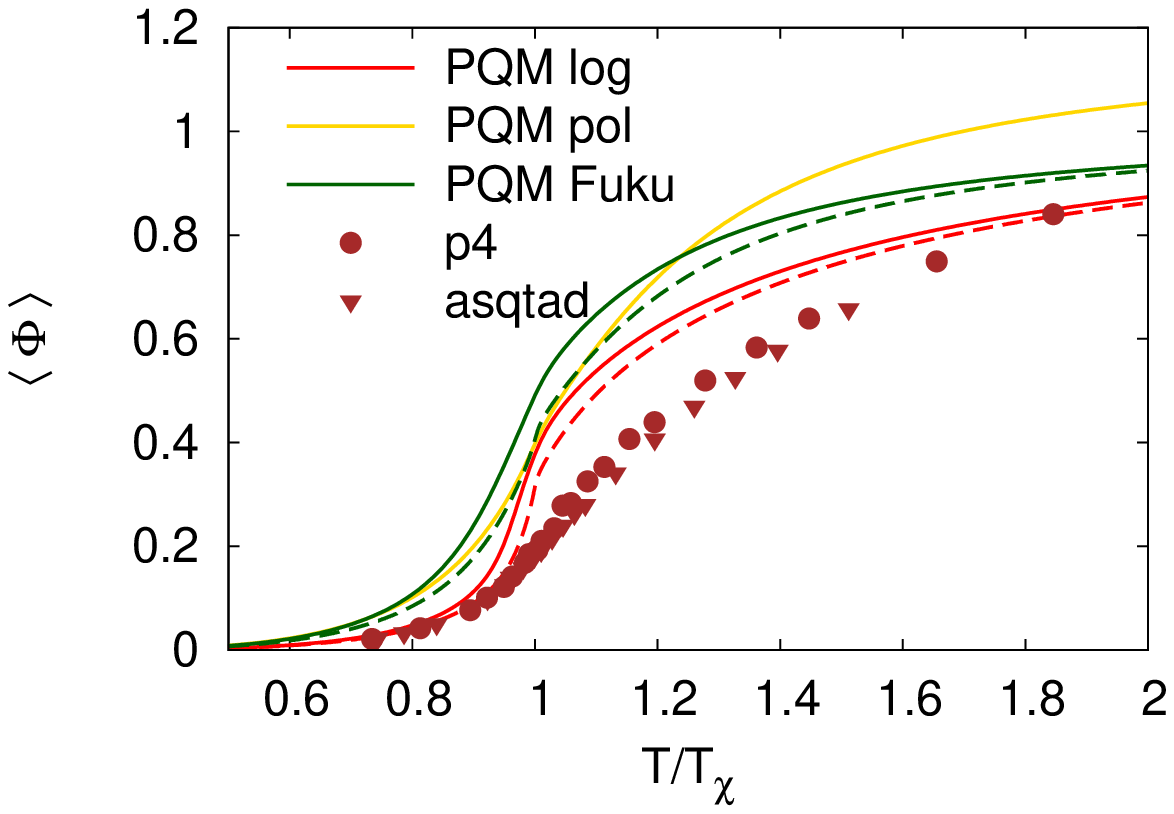}}
    \subfigure[$\ $Subtracted condensate $\Delta_{q,s}$]{\label{sfig:pqmdeltals}
    \includegraphics[width=\twofigs]{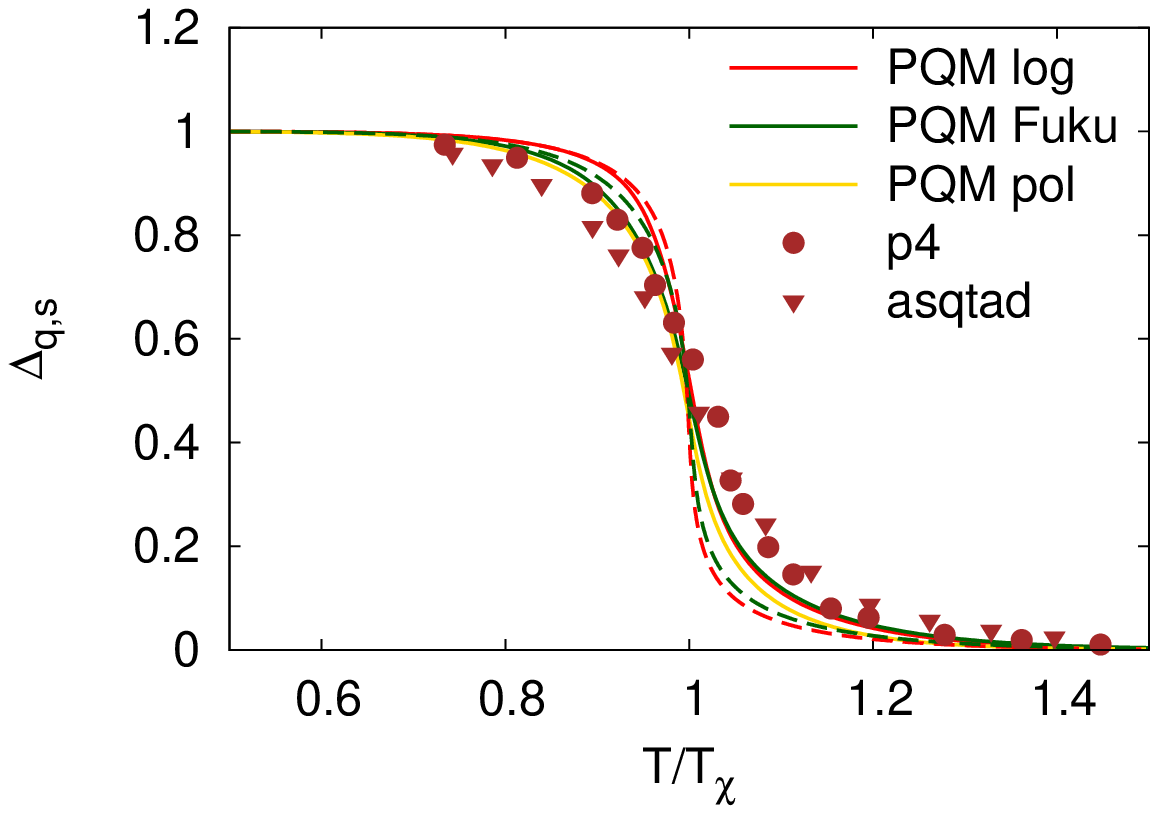}}
  \caption{\label{fig:pqmpolsigmpi}The Polyakov loop expectation value
    $\vev \Phi$ (left panel) and the subtracted condensate
    $\Delta_{q,s}$ (right panel) as a function of temperature for
    different Polyakov-loop potentials in comparison to lattice data
    for $N_\tau=8$ (symbols)~\cite{Bazavov:2009zn}. Solid lines
    correspond to larger pion and kaon masses as used in the lattice
    simulations, dashed lines to physical masses. }
\end{figure*}

In the following we have chosen a parameter set where both transitions
coincide at $\mu=0$ as suggested by lattice data which we use for comparison~\cite{Bazavov:2009zn}. This is the case for $m_\sigma=600\MeV$ and
$T_0=270\MeV$ where all three Polyakov-loop potentials yield
coinciding transitions.

\section{QCD thermodynamics}

In \Fig{sfig:pqmpoly} the Polyakov loop expectation value $\vev \Phi$
is shown as a function of temperature for three different Polyakov
loop potentials. For a proper comparison of the model results with the
HotQCD~\cite{Bazavov:2009zn} and RBC-Bielefeld lattice
data~\cite{Cheng}, we also adjust
the pion and kaon masses accordingly to $m_\pi=220 \MeV$ and
$m_K = 503\MeV$. Heavier meson masses yield also slightly heavier
constituent quark masses. The increased meson masses shift the
transitions also to higher temperatures. The strongest shift is seen
in the non-strange chiral transition since the quark masses are
largest in this sector and both transitions, the non-strange and
deconfinement one, still coincide. The strange sector is almost
unaffected. Solid lines in \Fig{sfig:pqmpoly} correspond to heavier
meson masses, used in lattice simulations and dashed lines to the
physical masses. The values are slightly higher for the Fukushima
potential than for the logarithmic potential which exhibits a sharper
crossover. The influence of the higher meson masses is mild. The
logarithmic potential comes closet to the lattice data in the broken
phase, while the difference between the model results and the lattice
is considerable in the restored phase for all potentials.

For $\mu=0$ the Polyakov loop expectation value $\vev \Phi$ is
available in model as well as in lattice calculations. This is in
contrast to the chiral condensate $\vev{\bar{q}q}$. A direct
comparison of $\vev{\bar{q}q}$ or $\vev{\bar{s}s}$ with lattice
simulations is difficult since unknown renormalization/normalization
factors are involved. Therefore, a better quantity to compare with
lattice simulations is the ratio
\begin{equation}
  \Delta_{q,s} = \frac{\langle \bar{q}q\rangle(T) -
    (\hat{m}_q / \hat{m}_s)\langle \bar{s}s\rangle(T)}{\langle
    \bar{q}q\rangle(0) - (\hat{m}_q / \hat{m}_s)\langle
    \bar{s}s\rangle(0)}
\end{equation}
which involves the non-strange $\langle \bar{q}q\rangle$ and strange
$\langle \bar{s}s\rangle$ condensates, correspondingly
$\vev{\sigma_{x}}$ and $\vev{\sigma_{y}}$ in our model calculation.
Since the bare quark masses $\hat{m}_q, \hat{m}_s$ are not directly
available, the ratios of the explicit symmetry breaking parameters of
our model are used instead, cf.~\cite{Meyer-Ortmanns:1994nt}.

The results for $\Delta_{q,s}$ are shown in \Fig{sfig:pqmdeltals}. The
model calculations are in better agreement than for the Polyakov loop
expectation value. However, also here, the lattice data exhibit a
smoother transition than the model results. The logarithmic potential
generates the sharpest transition while the Fukushima and polynomial
potential are closer to the lattice results. The influence of the
larger meson masses is stronger in the broken phase where the
transition becomes smoother for larger meson masses.

\begin{figure*}
\centering
\includegraphics[width=\twofigs]{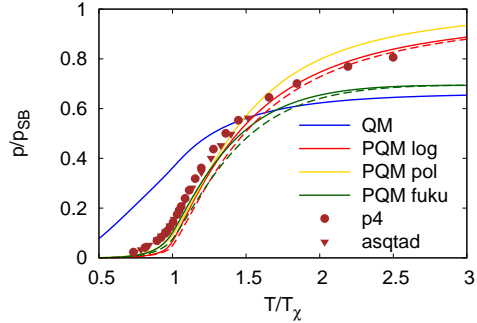}
\caption{\label{fig:pqmeos}The normalized pressure as a function of
  temperature. The model calculations (PQM model with various Polyakov
  loop potentials and the QM model \cite{Schaefer:2008hk}) are
  compared to lattice data similar as in Fig.~1.}
\end{figure*}

The pressure $p(T,\mu) = - \Omega\left(T,\mu \right)$ is directly
obtained from the thermodynamic potential with the normalization
$p(0,0) = 0$. In \Fig{fig:pqmeos} the (P)QM model pressures,
normalized to the SB value of the PQM model, are compared to lattice
data. As expected, the QM model \cite{Schaefer:2008hk} fails in
describing the lattice data for all temperatures, while the PQM model
results are in reasonable agreement with the data. The best agreement
with the lattice data is achieved with Fukushima's potential but fails
for temperatures above $1.5 T_\chi$. Physical meson (quark) masses
(dashed lines) push the pressure to lower values, in particular around
the transition.

\subsection{Taylor expansion and a novel algorithmic differentiation
  method}

As already mentioned one approach to overcome the sign problem in
finite density QCD lattice simulations is based on an extrapolation
from zero chemical potentials by a Taylor expansion. For the pressure
the expansion in powers of $(\mu/T)$ at $\mu=0$ reads

\begin{equation}
  \frac{p(\mu/T)}{T^4} = \sum_{n=0}^\infty c_n(T) \muT^n \quad
  \text{with} \quad	c_n(T) = \left.\frac{1}{n!}
    \frac{\partial^n\left(p(T,\mu) /T^4 \right)}{\partial
      \left(\mu/T\right)^n} \right|_{\mu=0}\ . 
\end{equation}
The coefficients $c_n(T)$ are accessible in QCD lattice simulations.
Because of the CP-symmetry of the QCD partition function, $Z(\mu) =
Z(-\mu)$, only even coefficients contribute. These have been
calculated by different lattice groups for $N_f=2$~\cite{Allton2005gk,
  Gavai} and $N_f=2+1$ quark flavors~\cite{Miao:2008sz}, currently up
to order $n \leq 8$. For the realistic $2+1$ flavor scenario
$N_\tau=4,6$ lattices have been used and higher coefficients still
have significant errors~\cite{Miao:2008sz}.

In principle, these coefficients can be obtained without any
limitations in model calculations. Doing this analytically is
 a tedious and difficult process. Furthermore, standard
numerical techniques like divided differences fail since the errors
increase rapidly. In order to proceed we have developed a novel
numerical technique based on \emph{algorithmic differentiation} (AD).
This technique allows the evaluation of higher derivatives to
extremely high precision. In fact, it is essentially limited only by
machine precision. Details of this method can be found in Ref.~\cite{ADTAylor}.

\begin{figure}
\centering
\includegraphics[width=.95\linewidth]{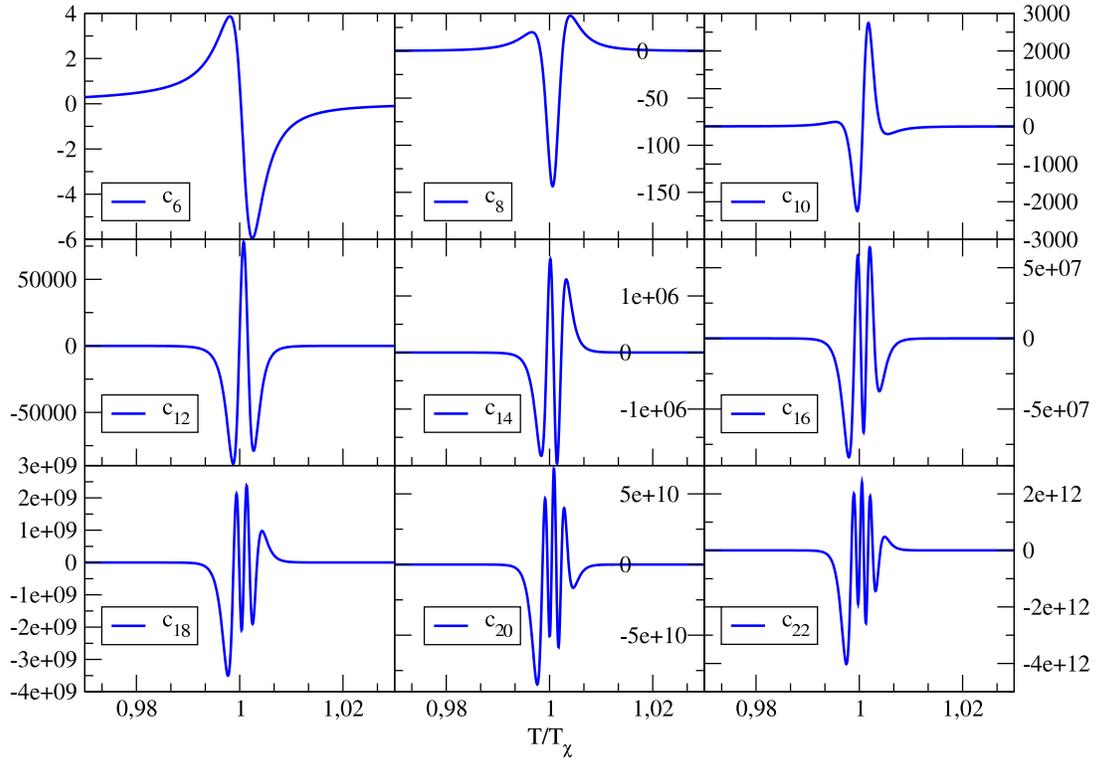}
\caption{\label{fig:taylorc}The Taylor coefficients $c_6$ to $c_{22}$
  in the PQM model with the logarithmic Polyakov-loop potential.}
\end{figure}

In \Fig{fig:taylorc} we show the resulting coefficients $c_6$ to
$c_{22}$ for the $2+1$ flavor PQM model with the logarithmic Polyakov
loop potential as a function of temperature. With increasing order
$n$, the coefficients $c_n$ start to oscillate around the transition
temperature $T_\chi$. Inside a narrow temperature window around
$T_\chi$, i.e., in the interval
$0.95 T_\chi \gtrsim T \gtrsim 1.05 T_\chi$ the oscillation amplitude
of the coefficients increases while they tend to zero far away of
$T_\chi$. This means that higher order coefficients become more and
more important in the expansion even for $\mu/T < 1$ and the
convergence of the series becomes questionable.
Nevertheless, the Taylor expansion might still be useful to gain some
information on the existence and location of a possible critical
endpoint (CEP) in the QCD phase diagram. For some recent estimates by
lattice groups using only a few coefficients see~\cite{Allton2005gk,
  Gavai}. However, the validity of these results
and the convergence of the series remain questionable.

An often considered quantity to extract some information on the CEP is
the quark number susceptibility
$\chi_q = (\partial^2 \Omega)/(\partial \mu^2)$. Since $\chi_q$
diverges exactly at the CEP it is a suitable quantity to locate this
point at least in model calculations. Its expansion can be expressed
with the same Taylor coefficients $c_n$ via
\begin{equation}
  \label{eq:taylorsuscep}
  \frac{\chi_q(\mu/T)}{T^2} = \sum_{n} {n (n-1) c_n(T)}
  \left(\frac{\mu}{T}\right)^{n-2}\ . 
\end{equation}

\begin{figure}
\centering
\subfigure[$n=4,8,12$, $\muT=0.8$]{\label{sfig:taylorsuscep08}
\includegraphics[width=\threefigs]{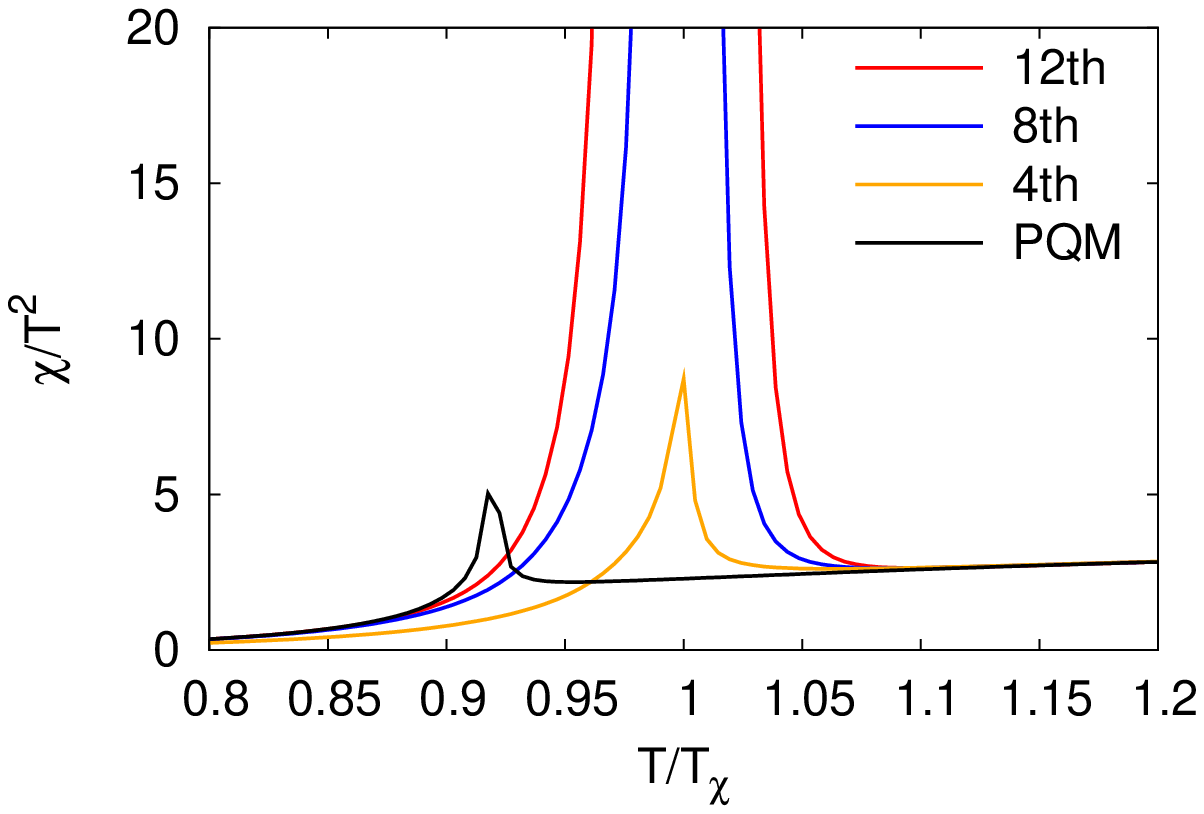}}
\subfigure[$n=4,8,12$, $\muT=\muT_c$]{\label{sfig:taylorsuscep09}
\includegraphics[width=\threefigs]{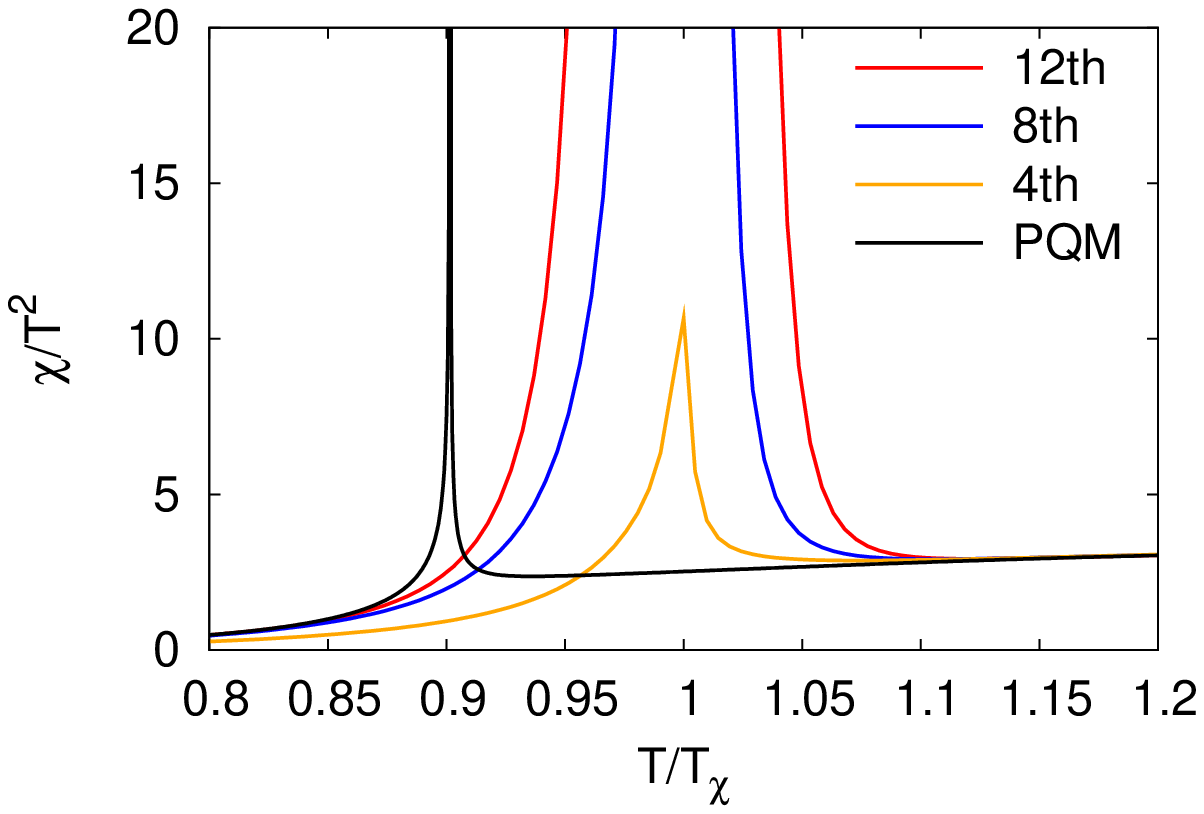}}
\subfigure[$n=4,8,12$, $\muT=1.0$]{\label{sfig:taylorsuscep10}
\includegraphics[width=\threefigs]{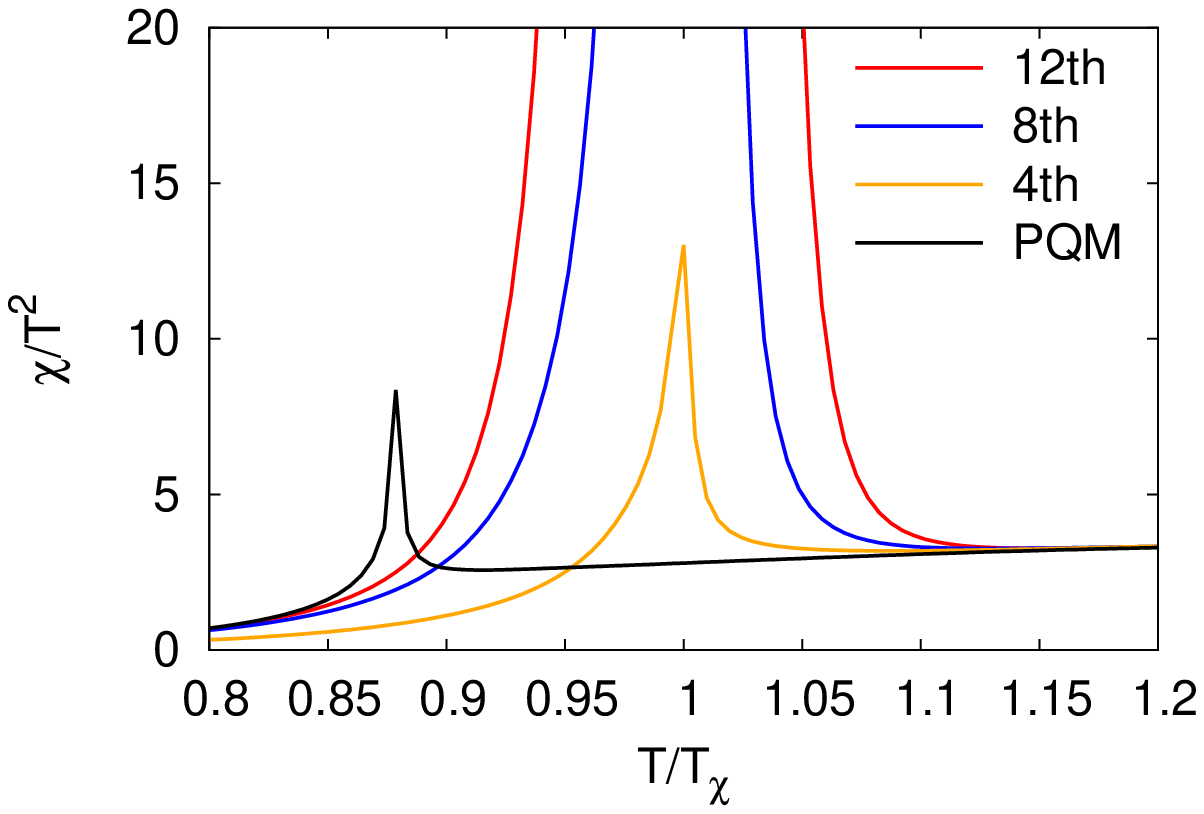}}
\caption{\label{fig:taylorsuscep}The quark-number susceptibility
  $\chi_q/T^2$ for different ratios of $\mu/T$ and different orders of
  the Taylor expansion. The black line, labeled with 'PQM', is the
  model calculation.}
\end{figure}

In \Fig{fig:taylorsuscep} we compare $\chi_q$ obtained with the Taylor
expansion of different orders with the evaluation of the model. For
the chosen parameters the CEP is located at $(T_c,\mu_c) \sim (185,
167)\MeV$, i.e., at $(\mu_c/T_c) \sim 0.9$ (middle panel). The other
two panels show $\chi_q$ for the ratios $(\mu/T)=0.8$ and $1.0$,
respectively. As expected, $\chi_q$ diverges at $T_c \sim 0.9 T_\chi$
and is finite for larger or smaller ratios $(\mu/T)$ near the critical
one (lines labeled with 'PQM'). Even the $12^\textrm{th}$ order
expansion neither reproduces this structure nor shows any significant
remnant of the divergence at the CEP. In the chirally broken phase,
i.e., on the left side of the peak, the agreement becomes better with
increasing orders of the Taylor expansion. The peak structure which is
visible in the Taylor expansion around $T \sim T_\chi$ is a pure
truncation artifact and a signal for its breakdown near the
transition. On the other hand, in the chirally restored phase at high
temperatures, the expansion again reproduces the model result, even
for $\mu/T=1$. This is in agreement with the high-temperature limit of
the coefficients $c_n$ since only the coefficients with $n\leq4$ have
a finite Stefan-Boltzmann value.

For a deeper understanding of the breakdown of the Taylor expansion it
is instructive to consider its convergence radius. It can be obtained
from the definition
\begin{equation}
  r = \lim_{n \rightarrow \infty} r_{2n} = \lim_{n \rightarrow \infty}
  \left|\frac{c_{2n}}{c_{2n+2}} \right|^{1/2}\ . 
\label{eq:convr}
\end{equation}
It is not known how well the radius $r$ is estimated by $r_n$ for a
finite value $n$.

\begin{figure*}[tb]
\centering
\subfigure[$n=4,8,12$]{
\includegraphics[width=\twofigs]{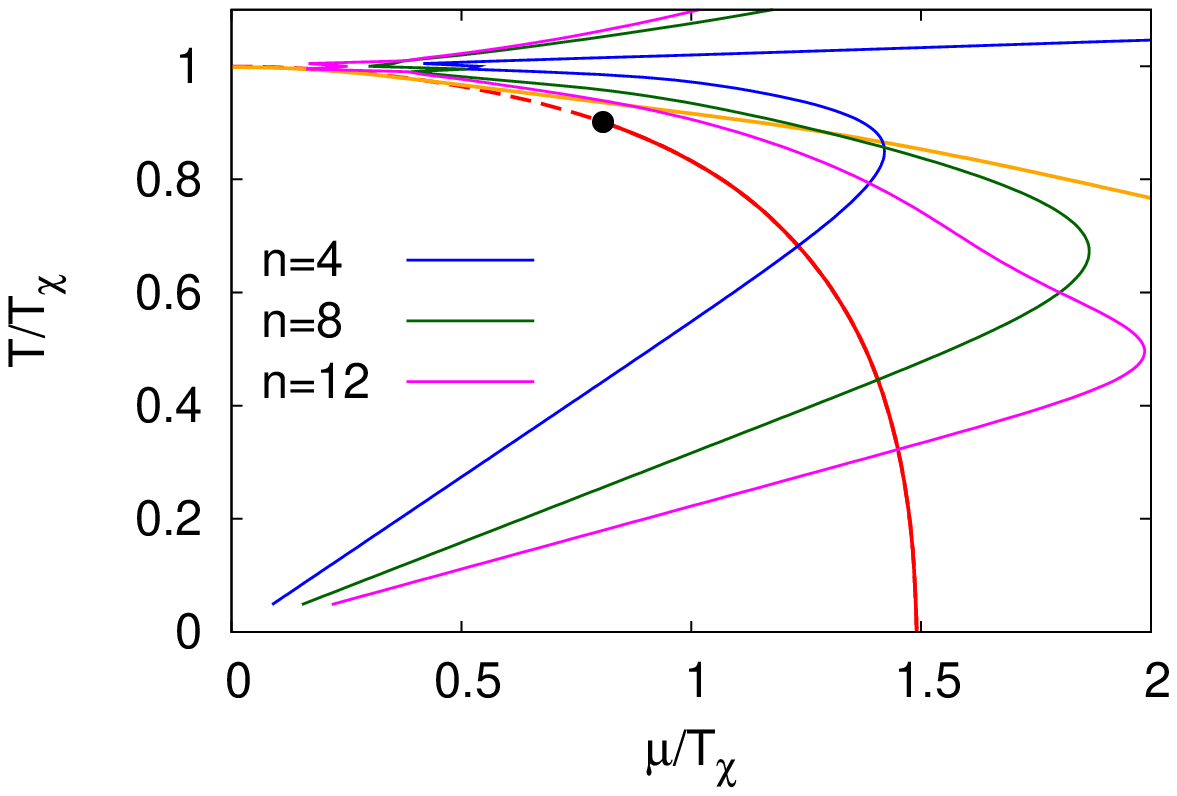}}
\subfigure[$n=16,20,24$]{
\includegraphics[width=\twofigs]{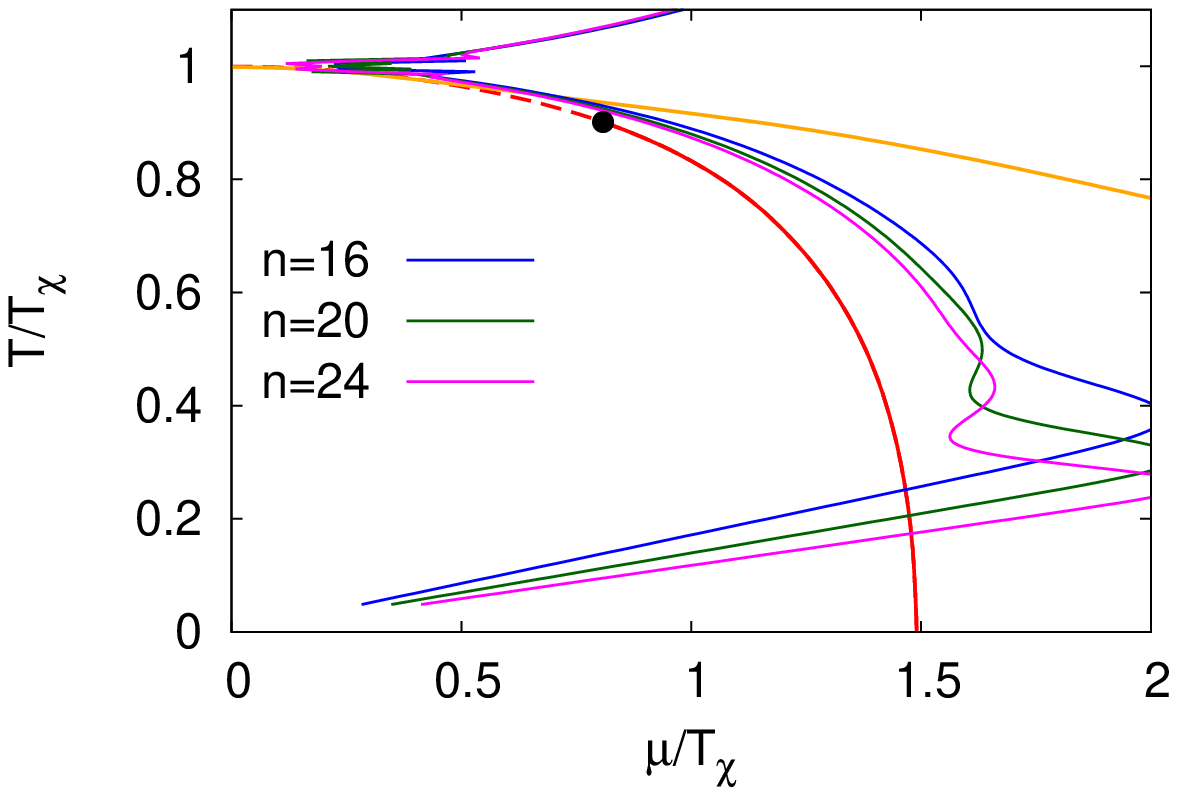}}
\caption{The convergence radii $r_n$ for different orders $n$ of the
  expansion in the PQM model with the logarithmic Polyakov loop
  potential ($T_0=270\MeV$). Also shown are the phase boundaries for
  the chiral transition (red line; dashed: crossover, solid: first
  order) and the deconfinement crossover transition (yellow line). The
  black dot indicates the CEP.}
\label{fig:convrad}
\end{figure*}

In \Fig{fig:convrad} we show the convergence radius for different
orders $n$ in the phase diagram together with the chiral and
deconfinement phase boundaries. For temperatures close to $T_\chi$ the
oscillations in the coefficients cause oscillations in $r_n$. For
smaller temperature, i.e., approaching the CEP from above, the CEP
lies inside the convergence region for lower $n$ values. For larger
$n$ values $r_n(T_c)$ approaches the CEP and points to $\mu_c$ once
$T_c$ is known. In the first order phase transition region, i.e., for
$T < T_c$ the interpretation of the convergence radius $r_n$ becomes
misleading. At a first order transition a new global minimum appears
in the grand potential and the order parameter jumps to the new
minimum. But the Taylor expansion is still performed around the
$\mu=0$ minimum and cannot capture the jump. This means that its
convergence radius is only valid for this $\mu=0$ minimum. Therefore
the results obtained with the expansion are only valid up to the first
order phase boundary.

Once $T_c$ is known, the convergence radius might provide an estimate
for the critical $\mu_c$ and to locate the CEP. In~\cite{Gavai}
the critical temperature was estimated by using the sign of the Taylor
coefficients. It contains information on the location of the CEP in
the complex chemical potential plane~\cite{Stephanov:2006dn}. For
temperatures below the critical one all coefficients should be
positive and the first root in the coefficients for $T<T_\chi$ should
determine $T_c$ in the limit $n \rightarrow \infty$. Since the location
of the CEP is precisely known in the model calculation we can verify
this idea. From the coefficients, shown in~\Fig{fig:taylorc}, we
observe that the first roots of the coefficients, starting at
$T\sim T_\chi$ for $n=6$, stay far away from the critical temperature
$T_c\sim 0.9 T_\chi$. From this one might conclude, that the Taylor
expansion yields a too large value for the critical temperature.
A more detailed study in this direction is underway
\cite{Taylorcoeff}.

\section{Summary}

In this talk we have presented recent work, which extends a two flavor
Polyakov-quark-meson model to $N_f = (2+1)$ quark flavors. The bulk
thermodynamic of this model in mean-field approximation is compared to
recent $(2+1)$-flavour QCD lattice simulations. The larger quark
masses on the lattice are also considered in the model calculations.
For the Polyakov-loop potential we examine three different ans\"atze
which all reproduce a first-order phase transition in the pure gauge
sector at the critical temperature $T_c = 270$ MeV. For the used
parameter sets of the Polyakov-loop potentials the chiral and the
deconfinement transition coincide at vanishing chemical potential.
With the Polyakov-loop a very good agreement of the QCD equation of
state in particular in the transition region up to temperature of $T
\sim 1.5 T_\chi$ is achieved in contrast to a pure three flavor
quark-meson model.

Furthermore, we analyze the finite chemical potential expansion of the
quark-number susceptibility in a Taylor series around vanishing
chemical potential. By means of a novel algorithmic differentiation
technique, we calculate the Taylor coefficients up to $24^\textrm{th}$
order in the PQM model for the first time. The knowledge of these
higher Taylor coefficients allows the systematic study of convergence
properties of the series. The peak in the quark-number susceptibility
$\chi_q$, seen in the Taylor expansion, is a pure artifact and related
to the breakdown of the expansion. In particular, the divergence of
$\chi_q$ at the critical endpoint in the phase diagram cannot be
captured by the Taylor series. Nevertheless, away from the phase
transition the Taylor expansion converges rapidly and coincides with
the model evaluation even for $\mu/T=1$. In the first-order region
discontinuities in the order parameter emerge which cannot be captured
by the Taylor series either. However, our analysis of the convergence
radius yields that the expansion might work even for $\mu/T>1$ if one
stays in the broken phase. But in lattice simulations this transition
boundary is not known. Thus, in order to make statements on the
possible location of a critical endpoint in the phase diagram with the
Taylor expansion technique our studies suggest that higher order
coefficients with $n>8$, more as currently available on the lattice,
are definitely required.

\paragraph{Acknowledgments}
The work of MW was supported by the Alliance Program of the Helmholtz
Association (HA216/ EMMI) and BMBF grants 06DA123 and 06DA9047I. JW
was supported in part by the Helmholtz International Center for FAIR.


%

\end{document}